\documentclass [preprint,
 amsmath,amssymb,
 aps,
 superscriptaddress,
12pt,floatfix,
]{revtex4-1}
\usepackage[title]{appendix}
\usepackage[T1]{fontenc}
\usepackage{lmodern}

\usepackage{graphicx}
\usepackage{capt-of}
\usepackage{geometry}
\usepackage{xcolor}
\usepackage{amsmath}
\usepackage{placeins}
\usepackage{array}

\usepackage{tabularray}

\usepackage{setspace}

\usepackage{url}
\usepackage{breakurl}

\begin{document}

\title{Apparent structural changes in contact patterns during COVID-19 were driven by survey design and long-term demographic trends}

\author{Tom Harris}
\affiliation{School of Computing and Information Systems, The University of Melbourne, Parkville, Victoria, Australia}

\author{Pavithra Jayasundara}
\affiliation{School of Public Health and Preventive Medicine, Monash University, Melbourne, Victoria, Australia.}

\author{Romain Ragonnet}
\affiliation{School of Public Health and Preventive Medicine, Monash University, Melbourne, Victoria, Australia.}

\author{James Trauer} 
\affiliation{School of Public Health and Preventive Medicine, Monash University, Melbourne, Victoria, Australia.}

\author{Nicholas Geard}
\affiliation{School of Computing and Information Systems, The University of Melbourne, Parkville, Victoria, Australia}

\author{Cameron Zachreson}
\affiliation{School of Computing and Information Systems, The University of Melbourne, Parkville, Victoria, Australia}

\date{\today}

\begin{abstract}
Social contact patterns are key drivers of infectious disease transmission. During the COVID-19 pandemic, differences between pre-COVID and COVID-era contact rates were widely attributed to non-pharmaceutical interventions such as lockdowns. However, the factors that drive changes in the distribution of contacts between different subpopulations remain poorly understood. Here, we present a clustering analysis of 33 contact matrices generated from surveys conducted before and during the COVID-19 pandemic, and analyse key features distinguishing their topological structures. While we expected to identify aspects of pandemic scenarios responsible for these features, our analysis demonstrates that they can be explained by differences in study design and long-term demographic trends. Our results caution against using survey data from different studies in counterfactual analysis of epidemic mitigation strategies. Doing so risks attributing differences stemming from methodological choices or long-term changes to the short-term effects of interventions. \\
\end{abstract}

\maketitle

\section{Introduction}
Infectious diseases can have devastating impacts on human populations and societies. For example, as of March 2024, the COVID-19 pandemic has resulted in over 775 million confirmed cases and over seven million deaths globally \cite{WHO_2024}. To limit the impact of epidemics, public health policy is designed to control infectious disease spread and reduce the associated health burden. Mathematical modeling of epidemic dynamics is used to guide policy design and implementation \cite{germann2022assessing,davies2020effects,di2020impact,zachreson2022covid}. To be able to accurately predict the scale, speed, and structure of epidemics, models must account for the factors that drive transmission. Such factors include the biological properties of the pathogen, risk factors in the host population, and crucially, the patterns of social contact mediating pathogen propagation between hosts \cite{wallinga2006using,edmunds1997mixes}.
 
Because of their key role in transmission of respiratory viruses such as Influenza and SARS-CoV-2, human contact patterns have been the subject of a sustained and intensifying global research effort \cite{Mosong2018,liu2021rapid}. The number of studies analysing close contact patterns relevant to the spread of respiratory pathogens has increased over the last few decades \cite{hoang2019systematic}. These studies have provided important insights facilitating the development of models that account for complex features of contact patterns such as age assortativity and variation in settings such as households, schools, and workplaces \cite{Mosong2018,gimma2022changes}.

In contact studies, data is often collected using surveys that ask participants to provide information on their recent close and casual contacts with others \cite{hoang2019systematic}. Survey data can be represented as a matrix that describes the contact rates within and between population strata. These contact matrices are used in transmission models to specify heterogeneous mixing rates between different types of individuals such as people of different ages \cite{van2013impact}. While it is recognised that contact patterns can vary substantially between jurisdictions, most surveys to date have been conducted in high-income countries (e.g., Belgium, Finland, Germany, Great Britain, Italy, Luxembourg, Netherlands and Poland \cite{Mosong2018}) with a small number conducted in low and middle-income countries (e.g., Vietnam \cite{Horby2011}, China \cite{Read2014}, Zimbabwe \cite{Melegaro2017} and Kenya \cite{Kiti2014}) \cite{hoang2019systematic}. 

Due to the expense and logistical complexity of survey studies, there is a continuing necessity to perform inference of contact patterns for populations where no survey data is available \cite{prem2017projecting}. However, the inference process increases the risk that contact patterns are misspecified in models. Misspecified models could substantially over- or underestimate policy-relevant projections (e.g., case load in at-risk age groups), with consequences for the calibration of public health interventions. To improve inference of contact patterns from available data and reduce the risks associated with misspecification, a systematic understanding of the mechanisms that drive variation in contact patterns over space and time is required \cite{hoang2019systematic,prem2017projecting,prem2021projecting,klepac2020contacts}.

Here, we identify features associated with variation in contact patterns by systematically comparing 33 different surveys conducted around the world over the last two decades. We observe that surveys naturally form into two main groups corresponding to studies performed before, and during the COVID-19 pandemic, respectively. By performing a detailed stratification of contact data, we demonstrate that these groupings primarily arise not from the behavioural changes which occurred during COVID-19 interventions, but from differing survey methodology and long-term demographic trends. The results of our analysis demonstrate the importance of accounting for survey methodology and timing when interpreting contact data for the development of disease transmission models.

\section{Results}
\subsection{Assessing contact matrix similarity}
\begin{figure}[h]
    \centering
    \includegraphics[width=\textwidth]{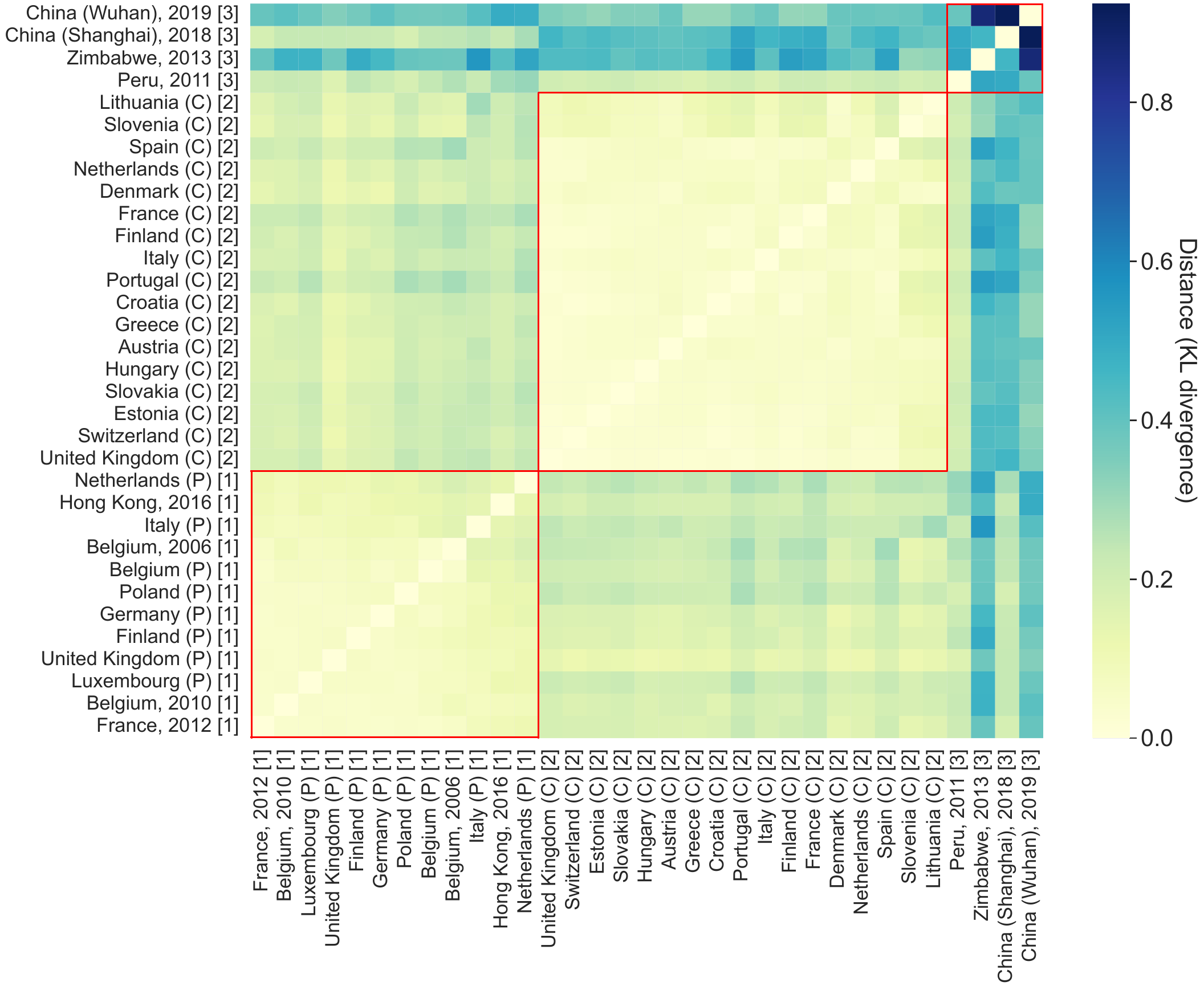}
    \caption{Pairwise comparison of contact matrices derived from different surveys, with distance expressed as the two-way Kullback-Liebler (KL) divergence. Three similar groups of contact matrices were found using spectral clustering (clusters in red) that corresponded to one group largely made-up of POLYMOD surveys (group 1), one group entirely comprised of CoMix (group 2) surveys, and a third group capturing other surveys analysed (group 3). Within-group similarity is high in both groups 1 and 2, and low in group 3, which captures outlier surveys in the analysed set. Surveys are listed by target country, whether they were conducted as part of the POLYMOD (P) or CoMix (C) studies, or if neither, the year they were conducted, with cluster identifier shown in brackets.}
    \label{fig:KL_full_set_n3}
\end{figure}

We compared contact matrices derived from 33 different contact surveys (8 conducted as part of the POLYMOD study in 2005-2006 and 17 conducted as part of the CoMix study in 2020-2022) \cite{Mosong2018,gimma2022changes}. Because our study focuses on identifying differences in structure, the contact matrix derived from each set of survey data was normalised to provide the relative rates of contact between people of different ages. We then calculated similarity using Kullback-Leibler (KL) divergence for each pair of normalised matrices. Based on pairwise similarity, we used spectral clustering to reveal three groups of matrices (Figure~\ref{fig:KL_full_set_n3}).

The first group is predominantly comprised of surveys conducted as part of the POLYMOD study. The second group corresponds to the entire set of CoMix surveys analysed (CoMix surveys from Belgium and Poland were excluded in our analysis). The third group is comprised of a smaller set of surveys that were not collected as a part of either the POLYMOD or CoMix studies. Notably, the POLYMOD and CoMix studies were performed under different conditions (e.g., CoMix conducted during the COVID-19 pandemic), and with different survey methods, but over approximately the same geographical region (Europe). The first two groups show a high level of within-group similarity. The third group has a comparably low within-group similarity, comprised of the set of outliers that do not fit into either of the first two groups.

\subsection{Analysis of contact patterns from the United Kingdom}
\begin{figure}
    \centering
    \includegraphics[width=\textwidth]{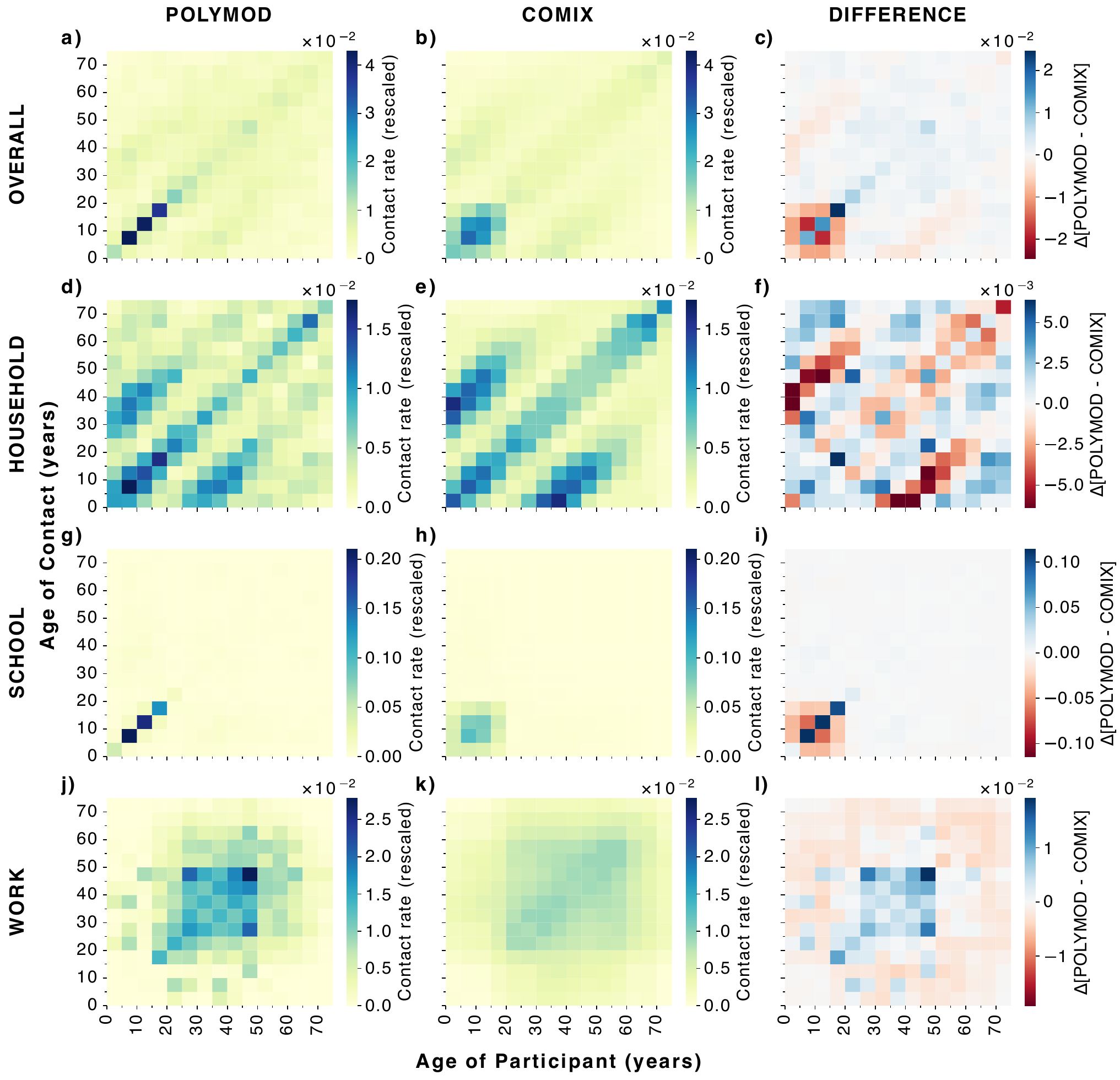}
    \caption{Detailed comparison of contact matrices derived from the POLYMOD and CoMix surveys conducted in the United Kingdom. The overall contact matrices (a,b) are derived from all contacts recorded in the UK POLYMOD and CoMix surveys, regardless of location. Contact matrices are also constructed from each survey filtered on the location where contacts occurred---household (d,e), school (g,h), workplace (j,k). Difference is computed for the overall matrices (c) and for the stratified matrices (f,i,l), by subtracting the CoMix contact matrix from the POLYMOD matrix. Note the eldest age group shown in each contact matrix captures all individuals greater than or equal to 70 years of age.}
    \label{fig:UK_poly_comix_by_context}
\end{figure}

To determine if variation in contact behaviour (due to e.g., nonpharmaceutical interventions) is responsible for the grouping in Figure~\ref{fig:KL_full_set_n3}, we compared the detailed contact patterns of the POLYMOD and CoMix surveys conducted in the United Kingdom (Figure~\ref{fig:UK_poly_comix_by_context}). We chose the UK surveys as representative examples due to their high within-group similarity in both groups 1 and 2 (see Supplementary Material \ref{mean_matrices}). Qualitatively, the contact patterns inferred from the POLYMOD survey (Figure~\ref{fig:UK_poly_comix_by_context}a) show a high level of assortative mixing, with a strong diagonal due to preferential contact between people close in age. There is a particularly high level of assortative mixing among children. The off-diagonal bands, common to many contact matrices previously studied, represent inter-generational mixing driven by children and parents within the same households. There is also a less intense, diffuse contact intensity between the age bands comprising working-age adults.

While the contact patterns inferred from the CoMix survey have a similar qualitative structure to those inferred from the POLYMOD survey (Figure~\ref{fig:UK_poly_comix_by_context}a~\&~b), quantitative comparison of the two contact matrices, stratified by the location where contacts occurred ---household, school, workplace--- revealed four distinct differences (Figure \ref{fig:UK_poly_comix_by_context}): 

\begin{enumerate}
\item{The CoMix patterns show a lower level of assortative mixing by age, especially in the younger age bands. Contact patterns associated with schools (Figure~\ref{fig:UK_poly_comix_by_context}g,h,i) reveal less assortative mixing among children in the CoMix patterns.}

\item{The POLYMOD patterns show a lower relative level of contact between children and older adults, stemming from differences in household contacts (Figure~\ref{fig:UK_poly_comix_by_context}d,e,f).}

\item{The CoMix patterns show a lower intensity of contact between working-age adults, demonstrated most clearly by contact patterns in workplaces (Figure~\ref{fig:UK_poly_comix_by_context}j,k,l).}

\item{The POLYMOD patterns show a higher intensity of contact between elderly individuals (i.e., 65+ years old) and children/middle-aged adults. This is most clearly observed in household contact patterns (Figure~\ref{fig:UK_poly_comix_by_context}d,e,f).}

\end{enumerate}

The POLYMOD and CoMix studies used different approaches to collecting data on the age of contacts. The POLYMOD study associates individual contacts with either exact ages or customised age brackets if the exact age was unknown to the survey participant \cite{Mosong2018}. The CoMix study asked participants about two different types of contact: group contacts (contacts reported in group settings, recorded with wider predefined age brackets) and individual contacts (recorded individually with narrower predefined age brackets) \cite{gimma2022changes}. 

\begin{figure}
    \centering
    \includegraphics[width=\textwidth]{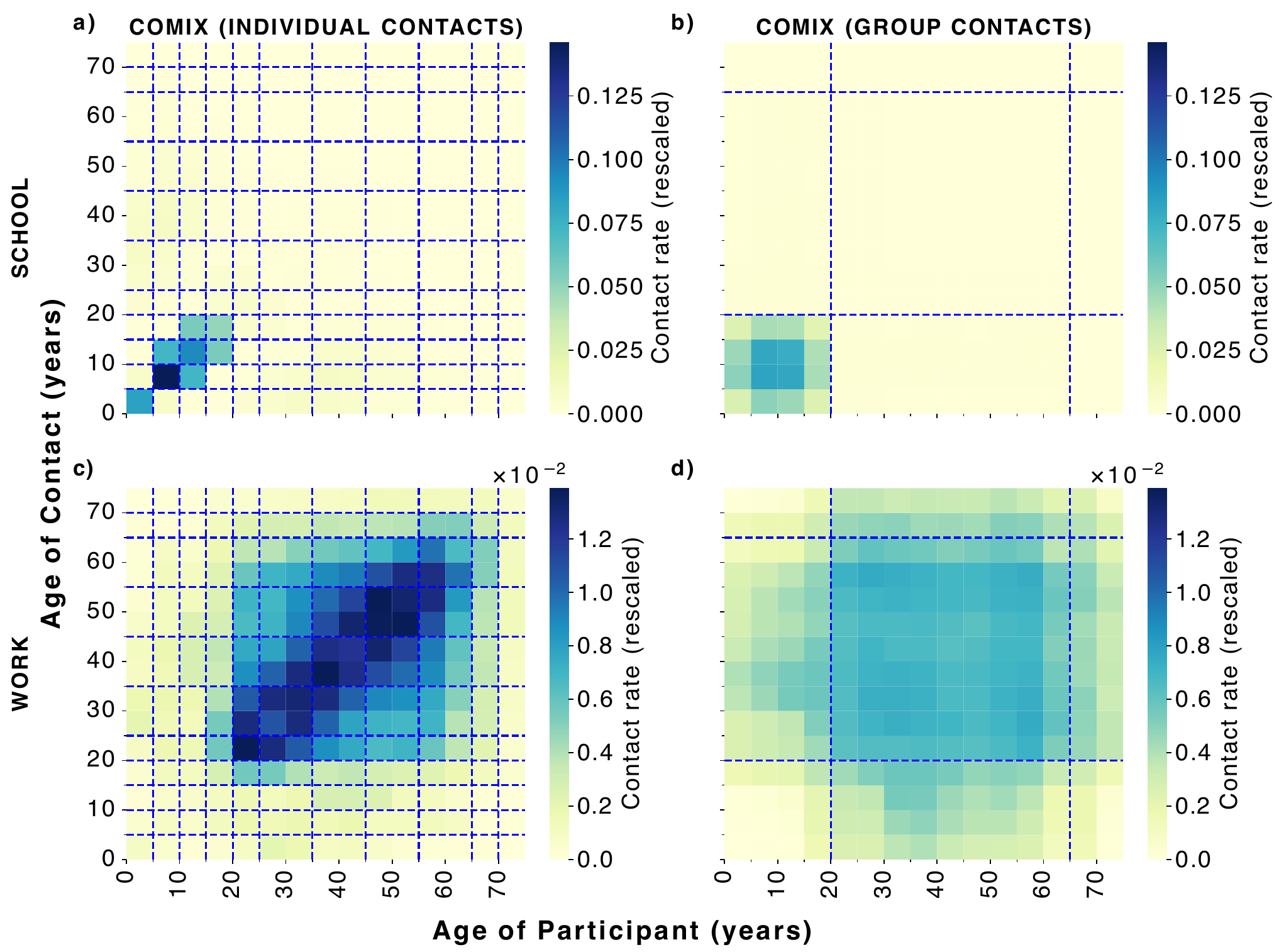}
    \caption{Contact matrices from the CoMix survey conducted in the United Kingdom, split on contact reporting method and location. Group contacts in the CoMix survey (contacts reported in group settings, recorded with wider predefined age brackets---b,d) are responsible for two of the observed structural differences between the UK CoMix and POLYMOD contact matrices. Namely, lower assortativity among children is related to group contacts in schools (b), and reduced contact among working age adults is related to group contacts in workplaces (d). Individual contacts in the CoMix survey (recorded individually with narrower age brackets---a,c) produce patterns more closely matching the POLYMOD contact patterns. The blue dashed lines approximately delineate the age brackets associated with individual and group contacts. Note the eldest age group shown in each contact matrix captures all individuals greater than or equal to 70 years of age.}
    \label{fig:UK_poly_comix_by_comix_contact_type}
\end{figure}

We examined the extent to which these different approaches affected the resulting contact matrices by separating individual and group contacts in the UK CoMix survey (Figure~\ref{fig:UK_poly_comix_by_comix_contact_type}). We found that patterns associated with individual contacts more closely resembled the POLYMOD patterns, suggesting the structural differences observed between surveys, in particular items (1) and (3) listed above, are explained in part by the inclusion of group contacts in the CoMix survey.

\begin{figure}
    \centering
    \includegraphics[width=\textwidth]{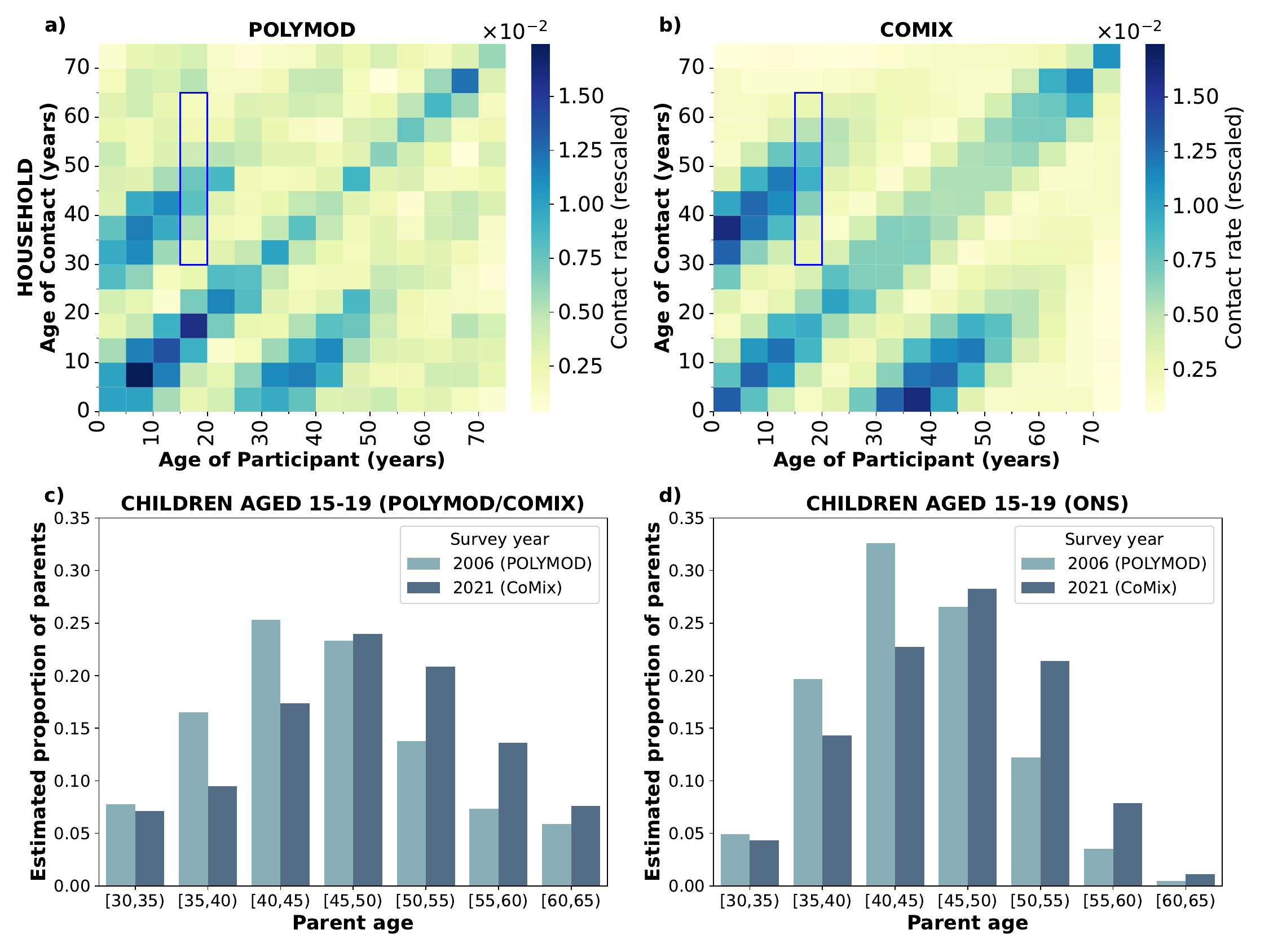}
    \caption{Similarities in the estimated change in parental age for children between 2006 and 2021, as approximated from the child-adult household contact patterns in POLYMOD and CoMix, and from age-specific fertility rates in the UK between 1985-2020. The shift in parental age of 15-19 year olds as estimated from age-specific fertility rates in the UK (d), aligns with the change in parental age estimated from the child-adult household contact patterns (c) in the POLYMOD (a) and CoMix (b) surveys. This similarity in parental age distribution suggests that the demographic shift in parental age contributes to differences observed between contact patterns derived from CoMix and POLYMOD surveys. The blue boxes highlight the age range considered likely to be parents in households that is used to derive an estimate of parental age (c). Note the eldest age group shown in each contact matrix captures all individuals greater than or equal to 70 years of age.}
    \label{fig:UK_poly_comix_ONS_parent_age}
\end{figure}

The higher intensity of mixing between children and older adults in household contact patterns reported in the CoMix survey could be explained by demographic changes occurring between the time of the POLYMOD study (2005-06) and the CoMix study (2020-2022). To examine how changes in parental age could be affecting household contact patterns over this time, we compared the age distribution of parents estimated from the child-adult household contact patterns measured in the UK POLYMOD and CoMix surveys, and from the UK age-specific fertility rate reported from the Office for National Statistics (ONS) between 1985-2020 (Figure~\ref{fig:UK_poly_comix_ONS_parent_age}) \cite{ons_2022}. Both distributions show a substantial increase in the typical age of parents between the time of the POLYMOD and CoMix surveys across all 5-year age brackets for children (see Supplementary Material \ref{demo_shift}), particularly in the 15-19 year old child bracket (Figure~\ref{fig:UK_poly_comix_ONS_parent_age}c~\&~d). The qualitative similarity suggests the increasing trend in parental age is likely a factor in the higher level of mixing between children and older adults in the CoMix survey.

\section{Discussion}
Clustering of contact matrices based on similarity revealed two distinct groupings associated with two of the largest contact survey studies completed to date: POLYMOD and CoMix. Notably, we do not see country level characteristics related to cultural, economic, or geographic factors emerge as drivers of matrix similarity. One would expect if these kind of factors were drivers of the underlying population contact patterns, as has been discussed elsewhere \cite{prem2017projecting,prem2021projecting}, countries with similar features, or even the same country surveyed on separate occasions, should emerge in robust groupings. We demonstrated in this study how survey methodology and demographic shift are likely reasons for the emergent clusters, which appeared to be more dominant drivers of matrix similarity than any country level characteristics in our analysis. The third cluster with low within-group similarity is composed of a number of surveys with unique characteristics that are likely responsible for their distinctive patterns. For example, the survey from Wuhan \cite{Zhang2020} was a retrospective survey with a substantial time difference (approximately 2 months) between when it was conducted and the time period participants were queried on.

Methodological differences between surveys complicates the comparison of contact patterns derived from them \cite{hoang2019systematic,gimma2022changes}. In our analysis of the POLYMOD and CoMix surveys from the UK, we demonstrated how the methodological choices in survey procedures can influence the resulting contact matrices. The approach taken by the CoMix survey to collecting data on group contacts resulted in a substantial number of contacts with very wide estimated age brackets (e.g., 18-64). While this approach to data collection may have eased the burden on survey participants, the resulting contact matrices created from these data were structurally different as a result of the inclusion of group contacts \cite{gimma2022changes,edmunds1997mixes}. For example, age assortativity among school contacts, a clear feature of the POLYMOD contact patterns, was largely lost with the inclusion of group contacts when processing the CoMix survey (Figure~\ref{fig:UK_poly_comix_by_comix_contact_type}a~\&~b).

Analyses of contact patterns measured during the COVID-19 pandemic have used pre-pandemic surveys (e.g., POLYMOD) as baseline counterfactuals to assess the impacts of policy interventions and endogenous behavioural changes \cite{gimma2022changes,liu2021rapid,wong2023social,jarvis2020quantifying,tizzani2023impact,coletti2020comix}. Cultural and demographic factors altering contact patterns between the POLYMOD study time period (2005-6) and the COVID-19 pre-pandemic time period should be considered when using POLYMOD data as a proxy for pre-pandemic behaviour \cite{jarvis2020quantifying,klepac2020contacts}. For example, changes in contact behaviour among teenagers since the POLYMOD study, thought to be related to rising online communication, have been factored into some pandemic contact behaviour analysis using POLYMOD data \cite{klepac2020contacts,jarvis2020quantifying}. Here, we demonstrated the effect of another long-term trend in contact behaviour between these two time periods: changes in age-specific fertility. Generally, the same increasing trend in age-specific fertility shown here for the United Kingdom has been observed in other countries, particularly in higher income countries \cite{world_fertility,OECD_2023}, which suggests this shift in household contact patterns is broadly relevant. Our results suggest changes to the distribution of parental ages should be factored into assessments of pandemic effects that use the POLYMOD data as a baseline and, more broadly, confirm the need to consider long-term demographic and cultural trends when interpreting older contact data \cite{klepac2020contacts,prem2021projecting}.

While we focus on parental age in our analysis, other demographic and cultural trends will affect contact patterns over time. For example, an increasing average age of individuals leaving the labour force, a trend observed in a number of countries over the last 20 years \cite{oecd_labour}, could lead to more older workers interacting in workplace settings. Our analysis of workplace contacts in the UK POLYMOD and CoMix data reflected this trend with a higher level of contact recorded in workplaces for the CoMix matrix being associated with older (55-70 years) individuals (Figure~\ref{fig:UK_poly_comix_by_context}).

The survey methods and demographic changes analysed here highlight two key considerations for users of contact data in infectious disease modelling. First, the methodology of a contact survey should be factored in to the interpretation of survey data. Understanding what information has been collected from survey participants and how this information should be interpreted is critical for ensuring the resulting disease models reasonably simulate the underlying contact patterns. For example, using the midpoint of a reported contact age range as an estimate of the specific age of a contact, as is done commonly when interpreting the POLYMOD surveys, could potentially be misleading when interpreting wide, predefined age brackets like those included in the CoMix surveys \cite{Mosong2018}. Second, differences in when a contact survey is conducted and the time period the survey data is intended to represent in a model should be addressed in model development and analysis. For example, including a sensitivity analysis that quantifies the robustness of model findings under alternate assumptions of the effect of long-term trends on old contact data (see, e.g., \cite{jarvis2020quantifying}). 

There is substantial evidence that people reduced social contact during the COVID-19 pandemic, both spontaneously and in response to non-pharmaceutical interventions \cite{gimma2022changes,wong2023social,liu2021rapid,wambua2023influence}. Here, we normalised contact rates to reveal similarities and differences in age-specific patterns of contact between different surveys, irrespective of the absolute magnitude of contact. Thus, our analysis reveals survey clusters arising from age-specific patterns of contact, rather than changes in magnitude resulting from COVID-19, clusters that we demonstrate can be attributed to survey methods and long-term demographic trends.

Interpretation of our results is subject to three key limitations of our methodology. First, while survey samples can provide useful insights into the contact patterns of a whole population, they remain approximations. While we apply common processing techniques (see Methods) to correct for known biases in contact survey data, there are potential sources of bias that these methods cannot address (e.g., recall bias, survey fatigue, sampling bias) \cite{hoang2019systematic,gimma2022changes}. Second, while we have attributed features distinguishing groups of contact patterns to methodological and demographic differences between surveys, we cannot entirely exclude the complex effects of the COVID-19 pandemic \cite{gimma2022changes,tizzani2023impact,Zhang2020}. The response to pandemic conditions, if common across the countries surveyed in the CoMix study, could be introducing structural features that influence the similarity between matrices (see Supplementary Material \ref{stringency}) for a more thorough analysis of this limitation). The third limitation arises from the interpolation method we chose when aligning survey data that reports age ranges for contacts (rather than defined ages). In our method we sample a contact's age from the reported range with weight derived from the underlying age demographics in the survey region (see Methods). While this approach improves upon uniform sampling by accounting for the age distribution of the population, it does not utilise information potentially present in the recorded characteristics of the participant. Employing refined methods for inference within large age brackets, such as a recent Bayesian approach \cite{dan2023estimating}, could improve this process.

Finally, another avenue for future work is to investigate if adjusting matrices for methodological differences in surveys (e.g., removing group contacts from the CoMix surveys) and known demographic shifts reveals underlying country level characteristics that drive matrix similarity. Any characteristics found to be associated with structural similarity in contact patterns could be used to improve matrix inference methodology.

In summary, our findings demonstrate that survey methodology and long-term demographic changes drive many observed differences between contact matrices. Age-specific fertility has introduced measurable changes in large-scale age-stratified contact rates, suggesting that fertility data could be used to improve extrapolation from historic patterns of contact. Our results highlight potential risks of using data from different studies when modelling counterfactual mitigation scenarios: doing so risks attributing differences stemming from methodological choices or long-term changes to the short-term effects of interventions.

\section{Methods}
\subsection{Selection and pre-processing of contact surveys}
\begin{figure}
    \centering
    \includegraphics[width = \textwidth]{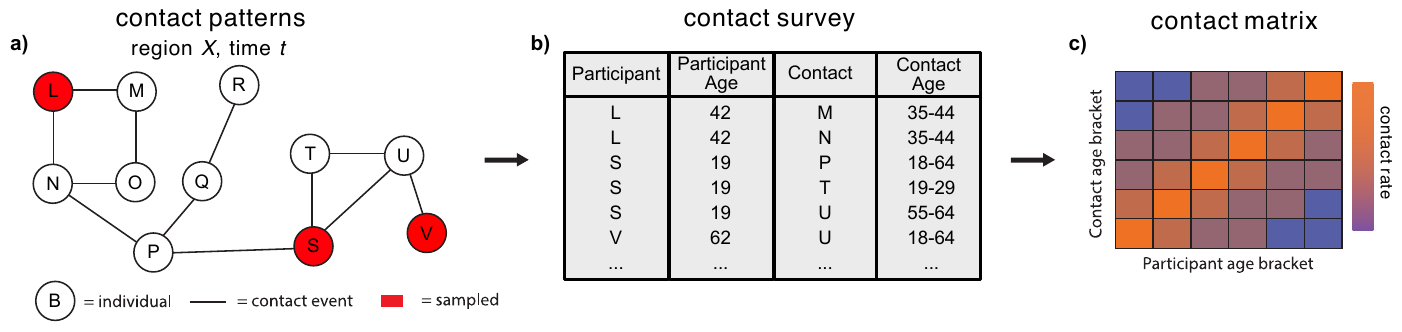}
    \caption{Schematic representation of contact patterns in some region \textit{X} at time \textit{t} (a), approximately measured with a contact survey (b) that is processed into a contact matrix (c). Participants in a contact survey report contact events with other individuals occurring over some survey time period. These surveys characterise contact patterns in particular target regions at specific points in time. Participants record their own age and the age of their contacts at varying levels of specificity, which can in part be related to the methodology of the survey. These contact surveys are processed into age-specific contact matrices representing contact rates between age groups which are used to inform infectious disease models.}
    \label{fig:schematic}
\end{figure}

Contact surveys approximate the contact patterns among individuals in a target region over some time period (Figure~\ref{fig:schematic}). To capture the age-related contact patterns between individuals, survey participants are asked to report their own age and estimate the ages of people they have had contact with either as exact ages, or using predefined or customisable age ranges, depending on the survey methodology. Survey data will reflect contextual factors associated with the target region and survey time period (e.g, whether a pandemic lockdown was in place), and factors related to the survey methodology (e.g., whether the survey questionnaire included pre-defined age ranges for participants to use when estimating contact ages). From survey data, contact matrices can be generated that extrapolate the patterns observed in the survey to the entire population that the survey was intended to represent (see, e.g., \cite{melegaro2011types}). A schematic illustration of the survey procedure and matrix construction is provided in Figure~\ref{fig:schematic}.

In this analysis, contact surveys were sourced from a social contact data sharing initiative, where an online collection of social contact data and analysis methods for infectious disease modellers are available (see Supplementary Material \ref{surveys}) \cite{willem2020socrates,Mosong2018,grijalva2015household,Melegaro2017,beraud2015french,Leung2017,zhang2019patterns,hens2009mining,willem2012nice,Zhang2020,tizzani2023impact,gimma2022changes,backer2023dynamics}. Surveys in this collection which only sampled participants of particular age ranges (e.g., studies of infant or child contact patterns), that did not provide sufficient information regarding the surveyed region, or did not include participants aged across all 5-year age bands used in our investigation were not considered in this analysis. For the survey from Wuhan, we used the pre-pandemic baseline survey from \cite{Zhang2020}. For surveys that were not nationally representative, we used region specific demographic data for processing of surveys from Peru \cite{grijalva2015household}, China \cite{zhang2019patterns,Zhang2020}, and Zimbabwe (filtered for peri-urban contact data only) \cite{Melegaro2017}. This demographic data were provided either in the study or sourced from \cite{Brinkhoff_2024}. Otherwise, we used the relevant national age demographics from the World Population Prospects \cite{wpp_2022}.

For the CoMix study, participants were asked to report the total number of group contacts they had in settings such as school and workplaces, for the age groups 0-17 years, 18-64 years, and 65+ years \cite{gimma2022changes}. These contacts were reported in addition to individual contacts for which participants record contacts and their characteristics individually using alternate, narrower age brackets (Under 1, 1-4, 5-9, 10-14, 15-19, 20-24, 25-34, 35-44, 45-54, 55-64, 65-69, 70-74, 75-79, 80-84, 85+). The individual and group contacts are both included in the generation of contact matrices from the CoMix data. In contrast, survey participants in the POLYMOD study provided either exact ages of contacts or customised age brackets if the exact age was unknown.

When an age range was provided for a participant or contact in a survey, we sampled a specific age within the relevant age range. For each surveyed population, we estimated a kernel density based on the age demographics of the country the survey was conducted in. We used this kernel to weight sampling within age ranges to more accurately capture the likelihood of someone being a particular age within an age range. For the POLYMOD surveys, when an exact contact age was unknown, the midpoint within the customised age ranges was provided; we used the midpoint in our analysis, as is typically done when interpreting POLYMOD data \cite{Mosong2018}.

We excluded participants without age information or who recorded contacts without age information. For all CoMix surveys, we censored all participants who recorded more than 50 contacts to 50 total contacts to remove the influence of the small number of participants who recorded large numbers of contacts, consistent with \cite{gimma2022changes}. We sampled 50 contacts from the set of recorded contacts, first selecting individual contacts, as they were more specific (i.e., narrower age bands), and then selecting a proportionate number of group contacts from workplace, school and other settings to reach 50 total contacts.

\subsection{Generating contact matrices from survey data}

To generate a contact matrix from survey data, we undertook the following steps:
\begin{enumerate}
    \item pre-processing to account for survey structure
    \item aggregation of survey responses
    \item re-weighting to account for survey bias
    \item correction for contact reciprocity
\end{enumerate}
We conducted a pre-processing step to ensure consistent analysis of surveys with different record structure. After which, we performed steps 2-4 above using \textit{socialmixr}. Once these steps are complete, an element of the resulting contact matrix $c_{ij}$ describes the rate of interaction between one individual in age group $i$, and the whole group of individuals in age group $j$ (where $i$ and $j$ are the row and column indices of the matrix, respectively). 

To motivate our analysis, we now consider the case in which we have survey information for some context $A$ and we would like to use it to infer contact patterns for some target context $B$, for which no direct survey exists, but for which the overall distribution of the population into age groups is known. We label the contact matrix derived from the survey of context $A$ as $c^A$, and we divide each element $c^A_{ij}$ by the total population of age group $j$ in context $A$: 
\begin{equation}
    c'^A_{ij} = \frac{c^A_{ij}}{N^A_j}\,, 
\end{equation}
where $N^A_j$ is the number of individuals in context $A$ who are in age group $j$. This process produces a {\it{per-capita}} contact matrix for context $A$ which we label $c'^{A}$. The per-capita contact matrix represents the information about contact structure as derived from the surveyed population. It is a symmetric matrix that expresses the affinity between individuals in the different sub-populations, independently of the number of individuals occupying those groups. It therefore serves as an ideal point of comparison when determining which pairs of surveyed populations will serve as effective proxies for other, target populations. If two populations have similar per-capita contact matrices, then re-scaling a contact matrix from one of these populations to the demographic distribution of the other will provide a close approximation, i.e., 
\begin{equation}
    c^{A\rightarrow B}_{ij} = c'^A_{ij} N^B_j \approx c^B_{ij}\,,
\end{equation}
where $c^{A\rightarrow B}$ is the inferred contact matrix for context $B$, and $c^B$ is the true contact matrix from context $B$ (which may not be available due to lack of survey data). 

We used the \textit{socialmixr} R package \cite{socialmixr}---a collection of methods for constructing age-structured contact matrices---to generate contact matrices for our analysis. See documentation published alongside this package for additional details on some of the methods used in this analysis \cite{socialmixr_intro}.

Given a contact survey and the underlying population age demographics, we calculated the per-capita contact matrix (see `Contact rates per capita' in \cite{socialmixr_intro}). In this matrix, element $i$,$j$ described the contact rate between an individual in age group $i$ and any specific individual in age group $j$. We corrected for contact symmetry (see `Symmetric contact matrices' in \cite{socialmixr_intro}) and aligned sample and population age distributions through weighting of participants (see `Participant weights' in \cite{socialmixr_intro}). To isolate contact patterns with particular characteristics (e.g., contacts occurring in schools), we used the `filter' method available in the \textit{socialmixr} package (see `Filtering' in \cite{socialmixr_intro}). When filtering on individual and group contacts for the UK CoMix survey, we used the contact age brackets recorded for each contact to determine the contact reporting method.

\subsection{Comparing rescaled, per-capita contact matrices}
We measured the similarity of two contact matrices by computing the two-way Kullback–Leibler (KL) divergence. The KL divergence ($K$) is a measure of how different one probability distribution $P$ is to another reference probability distribution $Q$, over some sample space $X$:
\begin{equation}
    K(P||Q) = \sum_{x \in X} P(x)\text{log} \left(\frac{P(x)}{Q(x)}\right)\,, 
\end{equation}
To compare two contact matrices $A$ and $B$ with $k$ total age groups, we first normalised each matrix to a probability distribution ($P^A$ \& $P^B$) describing the likelihood that the individuals in a contact event are in age groups $i$ and $j$:
\begin{equation}
    P^A_{ij} = \frac{1}{\sum^{k}_{r=0}\sum^{k}_{v=0}A_{rv}} \times A_{ij}\,, 
\end{equation}
\begin{equation}
    P^B_{ij} = \frac{1}{\sum^{k}_{r=0}\sum^{k}_{v=0}B_{rv}} \times B_{ij}\,, 
\end{equation}
After which, we computed the two-way KL divergence ($D$) between $P^A$ and $P^B$. We calculated the divergence with both $P^A$ and $P^B$ as the reference distribution ($Q$), and computed the mean of the two resulting values:
\begin{equation}
    K(P^A||P^B) = \sum_{x \in X} P^A(x)\text{log} \left(\frac{P^A(x)}{P^B(x)}\right)\,, 
\end{equation}
\begin{equation}
    K(P^B||P^A) = \sum_{x \in X} P^B(x)\text{log} \left(\frac{P^B(x)}{P^A(x)}\right)\,, 
\end{equation}
\begin{equation}
    D(P^A,P^B) = D(P^B,P^A) = \frac{K(P^A||P^B)+K(P^B||P^A)}{2}\,, 
\end{equation}
where the sample space $X$ refers to the set of matrix elements in $P^A$ and $P^B$. We used the {\it philentropy} R package to compute the KL divergence with an $\epsilon$ value of $10^{-9}$ to replace zeros in the empirical distributions. A schematic of our analysis workflow is shown in Supplementary Material \ref{matrix_comparison}.

\subsection{Spectral clustering}
We used spectral clustering to identify groups of surveys with similar contact patterns. Spectral clustering utilises a graph Laplacian of a similarity matrix---a measure of how similar elements in the data set are to one another---to divide elements into several groups \cite{von2007tutorial}. From the eigenvectors of the Laplacian, spectral clustering classifies elements based on their similarity to other elements in the data set. 

We measured the similarity between contact matrices, derived from our set of contact surveys, using the two-way KL divergence (see above). From the pairwise similarity scores between contact matrices, we used a k-nearest neighbour approach to construct a similarity matrix, which connects graph elements (in this case contact matrices) to the $k$ most similar elements in the dataset and weights the edges based on the pairwise similarity of the endpoints \cite{von2007tutorial}. From this similarity matrix, the graph Laplacian was computed and surveys classified into clusters \cite{von2007tutorial}.

We used the {\it kknn} R package to perform spectral clustering. Key inputs to the clustering method provided in this package are the number of clusters to estimate ($N_c$) and the number of neighbours ($N_n$) considered in the k-nearest-neighbours similarity graph. We varied these parameters to explore a wide variety of potential clustering alternatives (see Supplementary Material \ref{spec_cluster}). For our main analysis, we chose $N_{c}=3$ and $N_{n}=5$.

As spectral clustering is a stochastic process, we assessed the robustness of the clustering reported in this study. To do so, we ran each clustering application 1000 times to assess the consistency of the resulting groups (see Supplementary Material \ref{spec_cluster}). We note here the parameters associated with the reported clustering result in this study achieved consistent groupings across all 1000 repetitions.

\subsection{Estimating parental age from age-specific fertility and contact patterns}
We estimated the distribution of parental ages in the United Kingdom for two time points relevant to the POLYMOD and CoMix contact surveys, 2006 and 2021. We used age-specific fertility data collected by the Office for National Statistics (ONS) between 1985 and 2020 to estimate the proportion of parents within five year age bands in 2006 and 2021 \cite{ons_2022}. The ONS data contains the number of births ($B^t_{a}$) in England and Wales (assumed here to be representative of the UK) each year $t$, split on the age of the mother $a$ using five year age bands. We assumed the total number of births reported for mothers in each five year age band (e.g., $B^{t}_{20-24}$ for 20-24 year old mothers) was equally distributed across each specific age (e.g., $B^{t}_{20} = B^{t}_{20-24} / 5$), and that the `under 20' and `40 and over' age groups reported in the ONS data could be represented by age groups 15-19 years and 40-44 years. We also assume the distribution of mothers' ages was a reasonable approximation of the distribution of parental (i.e., mothers and fathers) ages. Using this data, we were able to derive an estimate of the number of parents ($F^{t}_{a,b}$) in age group $a$ for a specific survey year $t$ that have a child within an age group $b$:
\begin{equation}
    F^{t}_{a,b} = \sum^{b_2}_{i=b_1}\sum^{a_2}_{j=a_1}B^{t-i}_{j-i}\,, 
\end{equation}
where $a_1$,$a_2$ and $b_1$,$b_2$ are the lower and upper limits of age groups $a$ and $b$, respectively. For example, to estimate the number of parents in age group 25-29 years in 2021 that have children aged 5-9 years, we computed:
\begin{equation}
    F^{2021}_{25-29,5-9} = B^{2016}_{20} + B^{2016}_{21} + B^{2016}_{22} + B^{2016}_{23} + B^{2016}_{24} + B^{2015}_{19} + B^{2015}_{20} + B^{2015}_{21} + ... + B^{2012}_{20}
\end{equation}
From these age group estimates of the total number of parents, we were able to compute the proportion of parents in 5 year age bands for each child age group (i.e., 0-4 years, 5-9 years, 10-14 years, 15-19 years). 

To compare to the estimated distribution derived from the ONS data, we estimated the distribution of parental age from the UK POLYMOD and CoMix household contact matrices. We selected the age specific contact rates between each child age bracket (i.e., 0-4 years, 5-9 years, 10-14 years, 15-19 years) and the five year age brackets between 15 and 45 years older than each child age group, which we defined as likely parental ages. For example, for the 15-19 years age group, we selected the household per-capita contact rates reported between 15-19 year old's and the 30-34 years, 35-39 years, 40-44 years, 45-49 years, 50-54 years, 55-59 years, and 60-64 years age groups  (see blue boxes in Figure \ref{fig:UK_poly_comix_ONS_parent_age}). For each set of per-capita contact rates, we divided each element by the sum of the set to produce an estimate of the proportion of parents in each age group.


\section{Data and materials Availability}
The data and analysis code are available from a GitHub repository: \url{https://github.com/tomharris4/contact-matrix-analysis-public}. Archived version is available at Zenodo \url{https://doi.org/10.5281/zenodo.11254978}.


\section{Author Contributions}
Conceptualization: T.H., P.J., R.R., J.T., N.G., and C.Z., Investigation: T.H., N.G., and C.Z., Methodology: T.H., P.J., R.R., J.T., N.G., and C.Z., software: T.H., writing—original draft: T.H., N.G., and C.Z., writing—review and editing: T.H., P.J., R.R., J.T., N.G., and C.Z. 

\section{Competing Interests}
The authors declare that they have no competing interests.

\section{Acknowledgments}
This work was supported by seed funding from the National Health and Medical Research Council Centre of Research Excellence in 2022 (SPECTRUM; NHMRC APP 1170960). Thomas Harris is supported by an Australian Government Research Training Program Scholarship and The University of Melbourne Elizabeth and Vernon Puzey Scholarship.

\section{References}
\bibliography{references.bib}

\newpage

\newcommand{\beginsupplement}{%

 \setcounter{table}{0}
   \renewcommand{\thetable}{S\arabic{table}}%
   
     \setcounter{figure}{0}
      \renewcommand{\thefigure}{S\arabic{figure}}%
      
      \setcounter{page}{1}
      \renewcommand{\thepage}{S\arabic{page}} 
      
      \setcounter{section}{0}
      \renewcommand{\thesection}{S\arabic{section}}
      
      \setcounter{equation}{0}
      \renewcommand{\theequation}{S\arabic{equation}}
     }

\beginsupplement

\FloatBarrier
{\bf \Large{Supporting Information for:\\ Apparent structural changes in contact patterns during COVID-19 were driven by survey design and long-term demographic trends}}

\section{Method for calculating mean contact matrices of Groups 1 \& 2}
\label{mean_matrices}
\begin{figure}[h]
    \centering
    \includegraphics[width = \textwidth]{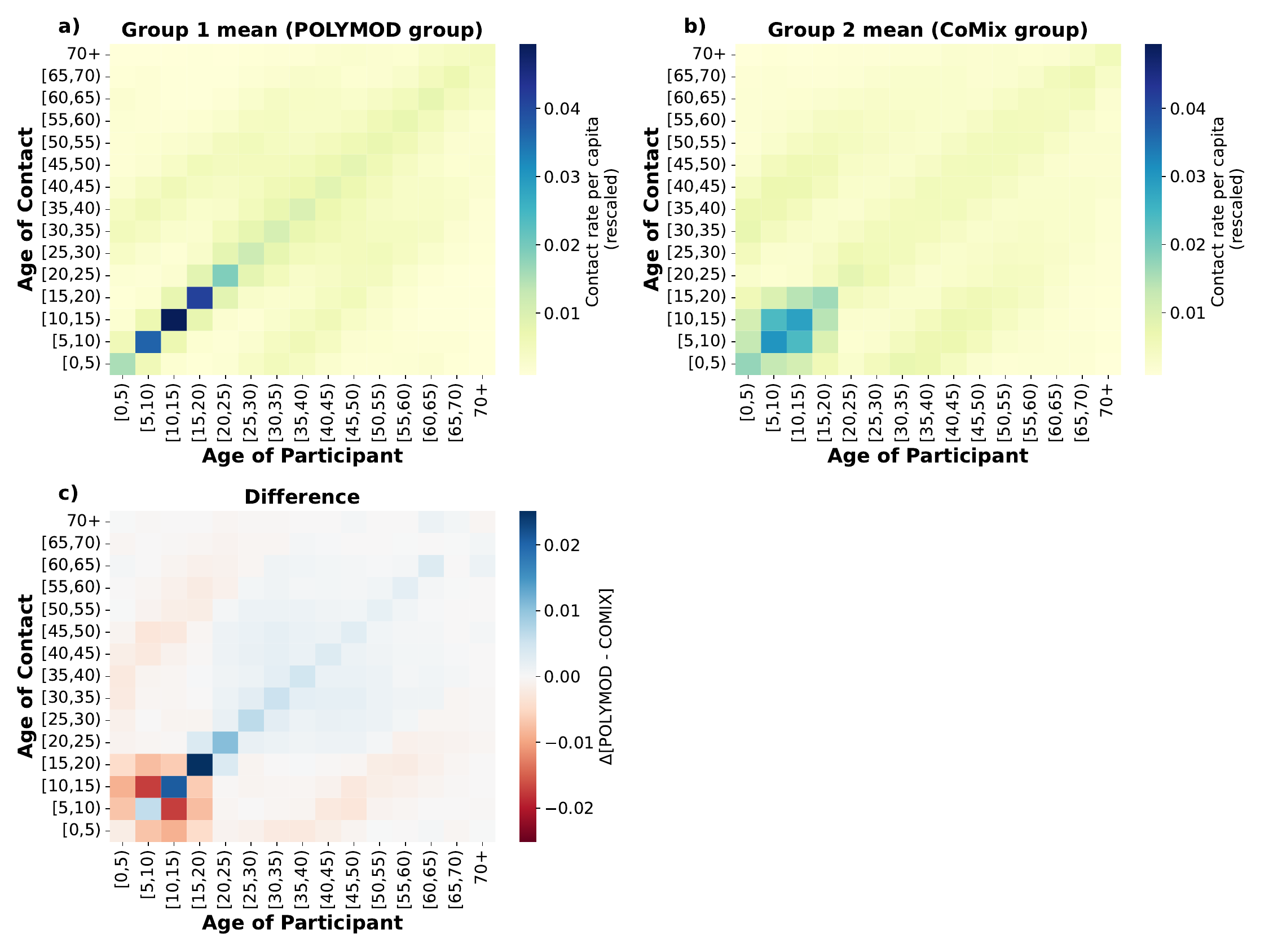}
    \caption{Comparison of the mean contact matrices derived from group 1 (a) and group 2 (b) of our clustered similarity matrix. The contact matrices are computed as the element-wise average for contact matrices in group 1 (POLYMOD group) and group 2 (CoMix group). Difference is the subtraction of the group 2 mean contact matrix from the group 1 mean contact matrix.}
    \label{fig:g1_g2}
\end{figure}

To establish whether the contact patterns derived from the United Kingdom surveys were representative of the contact patterns computed across groups 1 (POLYMOD) and 2 (CoMix) of our clustered similarity matrix, we computed the mean contact matrices for both groups (Figure~\ref{fig:g1_g2}). From our re-scaled, per-capita contact matrices in each group, we calculated the element-wise average for contact matrices in group 1 ($c^1$) and group 2 ($c^2$):
\begin{equation}
    c^1_{ij} = \frac{\sum_{c \in C^{1}}c_{ij}}{|C^1|}\,, 
\end{equation}
\begin{equation}
    c^2_{ij} = \frac{\sum_{c \in C^{2}}c_{ij}}{|C^2|}\,, 
\end{equation}
where $C^{1}$ and $C^{2}$ represent the set of contact matrices in groups 1 and 2, respectively.

The mean matrices for groups 1 and 2 is shown in Figure \ref{fig:g1_g2}. The matrices reflect similar contact patterns to those found in the UK POLYMOD and CoMix contact matrices (see Main Text, Figure~2). Specifically, when we compare the CoMix group mean matrix with the POLYMOD group mean matrix, we see the CoMix group mean matrix is less assortative, has increased mixing among older adults and children, and less mixing among working-age adults. The similarity of these group mean matrices to the UK contact matrices suggests the contact patterns found in the UK contact matrices are generally representative of the patterns captured in contact matrices within these groups.

\newpage
\newpage

\section{Impact of demographic shift in parental age - all child age groups (0-20 years)}
\label{demo_shift}
\begin{figure}
    \centering
    \includegraphics[width = \textwidth]{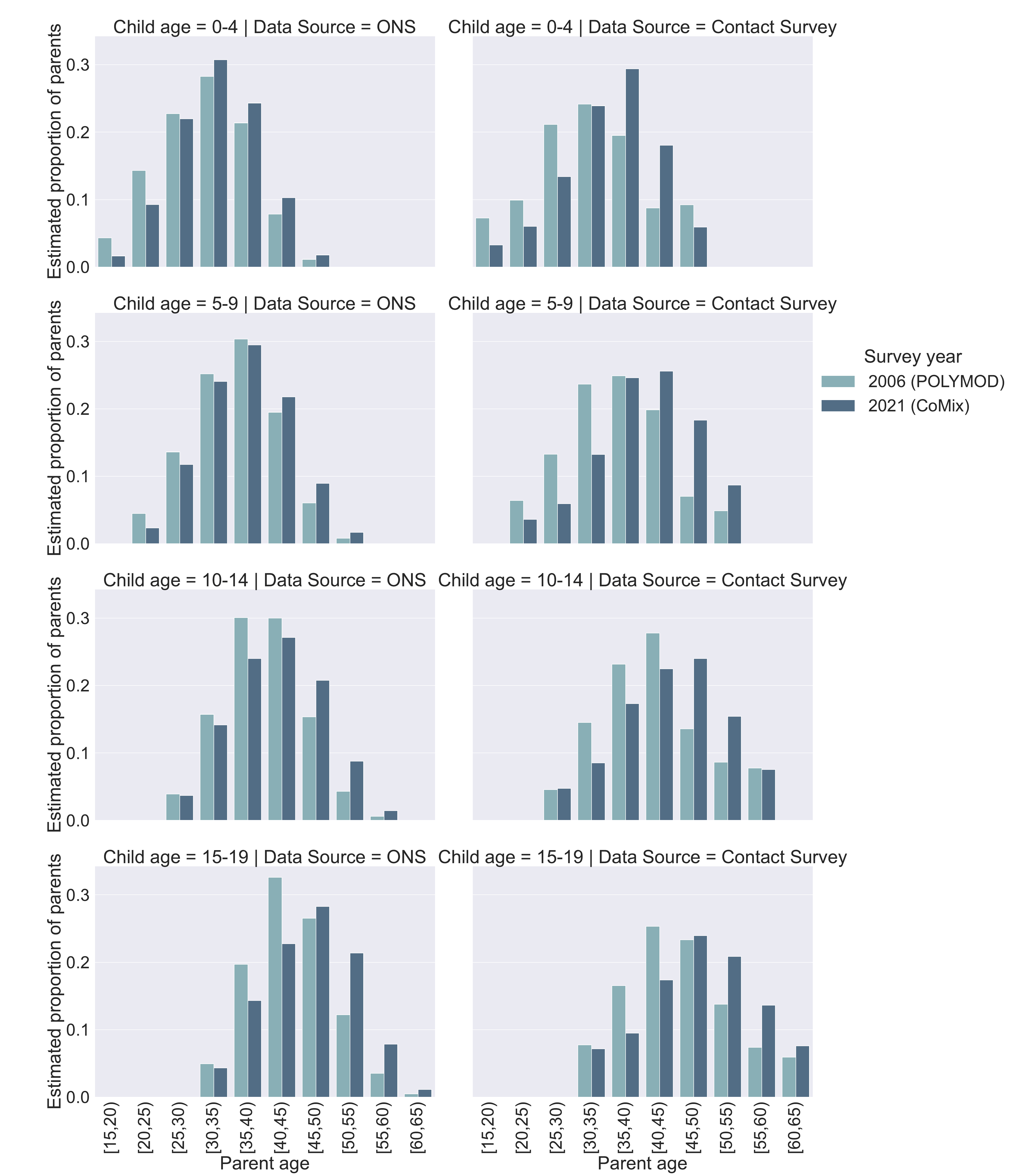}
    \caption{Similarities in the estimated change in parental age between 2006 and 2021, as approximated from the child-adult household contact patterns in POLYMOD and CoMix, and from age-specific fertility rates in the UK between 1985-2020. The shift in parental age of children as estimated from age-specific fertility rates in the UK, aligns with the change in parental age estimated from the child-adult household contact patterns in the POLYMOD and CoMix surveys. This similarity in parental age distribution suggests that a demographic shift in parental age contributes to differences observed between contact patterns derived from CoMix and POLYMOD surveys.}
    \label{fig:full_parents}
\end{figure}

We compared the estimated age distribution of parents for all child age groups (0-4 years, 5-9 years, 10-14 years, 15-19 years), computed separately using the age-specific fertility rates in the UK recorded by the Office for National Statistics (ONS) \cite{ons_2022}, and the household contact patterns captured in the POLYMOD and CoMix surveys conducted in the UK (Figure \ref{fig:full_parents}). Across all child age groups, estimates generally reflected an increasing parental age between the POLYMOD survey year (2006) and the CoMix survey year (2021). There is a qualitative similarity between the parental age estimates generated from the ONS data and the POLYMOD and CoMix contact surveys, suggesting long-term changes in parental age is a factor in the changes to contact patterns between the POLYMOD and CoMix surveys.

\newpage
\section{Impact of intervention stringency on contact patterns derived from the UK CoMix survey}
\label{stringency}
To better understand the effect of nonpharmaceutical interventions on contact patterns recorded during the COVID-19 pandemic, we split the UK CoMix survey based on the stringency of the interventions applied on each survey day. We derived contact matrices from each sub-sample of the survey to capture how the contact patterns differed under varying levels of intervention intensity. To quantify stringency, we used the Oxford Coronavirus Government Response Tracker (OxCGRT) project \cite{owidcoronavirus,hale2021global}. They have computed a stringency index over time for a range of countries during the COVID-19 pandemic. The stringency index is a composite measure based on nine response indicators, such as school/work closures and travel bans. On a given day, the index takes a value between 0 and 100, with 100 indicating the strictest setting.

\begin{figure}
    \centering
    \includegraphics[width = \textwidth]{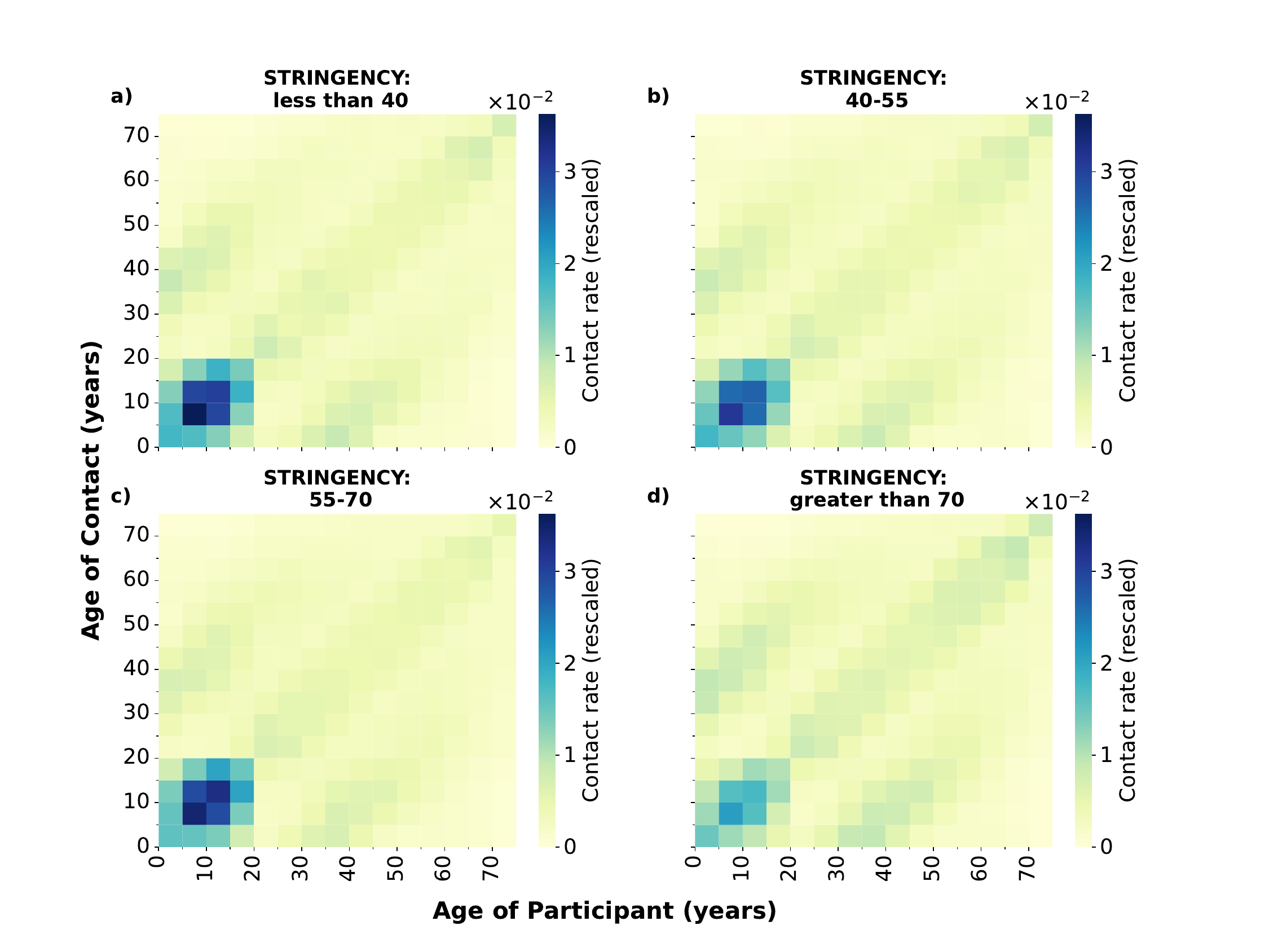}
    \caption{Overall contact matrices derived from CoMix survey conducted in the United Kingdom, split on stringency of interventions on survey days. The stringency index is a composite measure based on nine response indicators, such as school/work closures and travel bans. On a given day, the index takes a value between 0 and 100, with 100 indicating the strictest setting. Contact patterns derived from survey data recorded on days with stringency levels less than 40, 40-55, 55-70, and 70+, are shown in panels a, b, c, and d, respectively.}
    \label{fig:stringency_overall}
\end{figure}

\begin{figure}
    \centering
    \includegraphics[width = \textwidth]{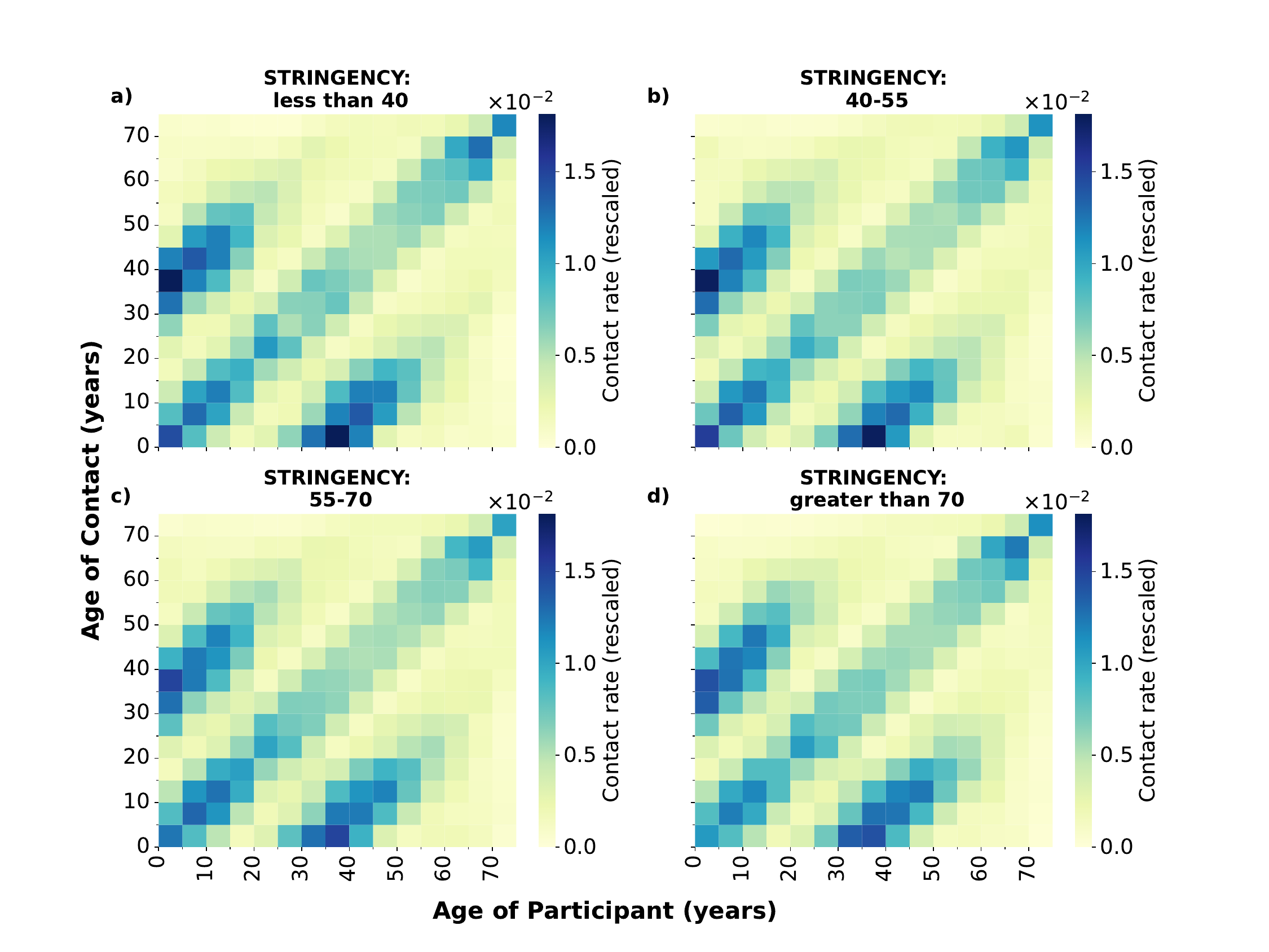}
    \caption{Household contact matrices derived from CoMix survey conducted in the United Kingdom, split on stringency of interventions on survey days. Stringency is quantified using a composite measure based on nine response indicators, such as school/work closures and travel bans. On a given day, the index takes a value between 0 and 100, with 100 indicating the strictest setting. Household contact patterns derived from survey data recorded on days with stringency levels less than 40, 40-55, 55-70, and 70+, are shown in panels a, b, c, and d, respectively.}
    \label{fig:stringency_household}
\end{figure}

\begin{figure}
    \centering
    \includegraphics[width = \textwidth]{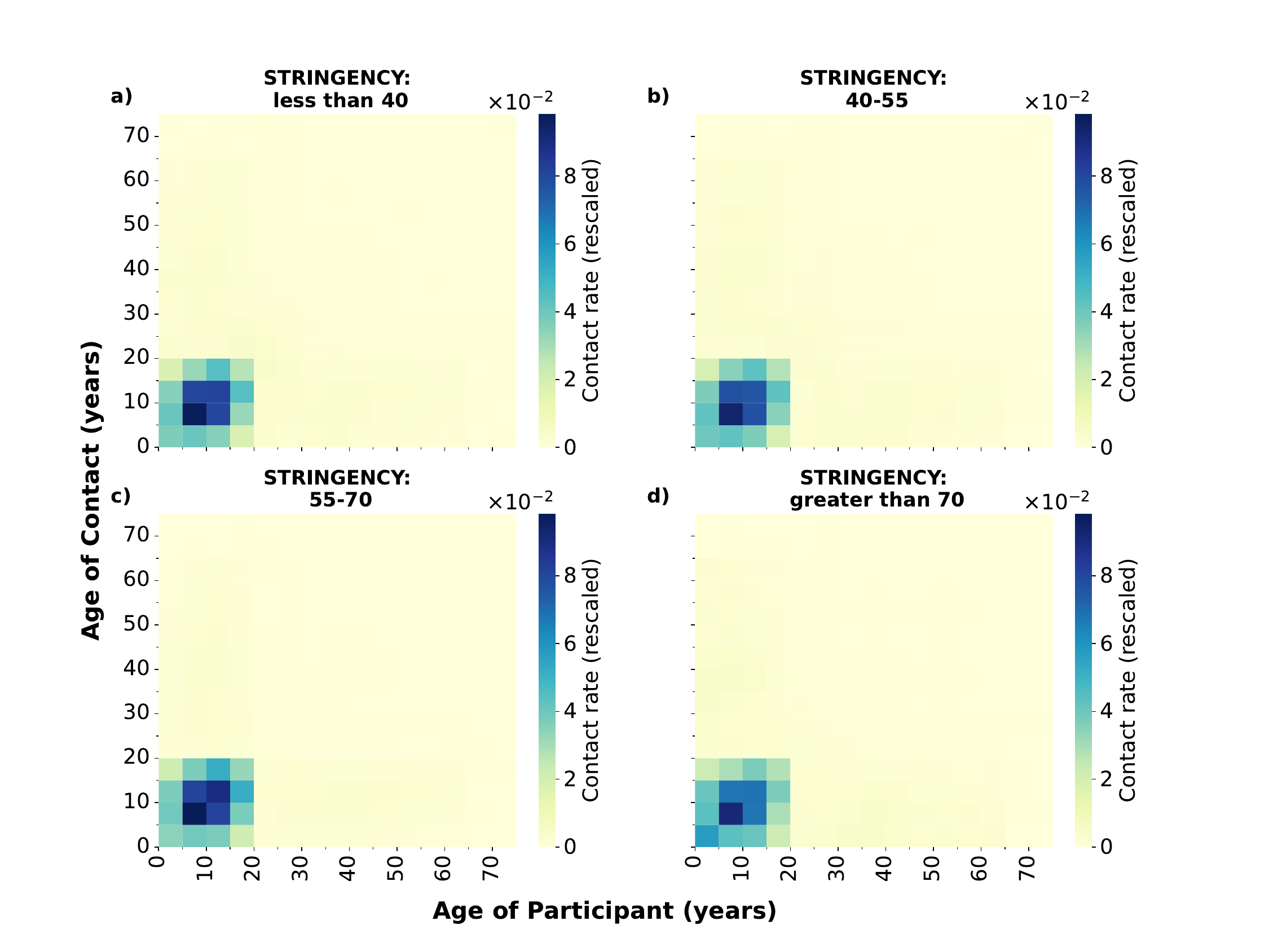}
    \caption{School contact matrices derived from CoMix survey conducted in the United Kingdom, split on stringency of interventions on survey days. Stringency is quantified using a composite measure based on nine response indicators, such as school/work closures and travel bans. On a given day, the index takes a value between 0 and 100, with 100 indicating the strictest setting. School contact patterns derived from survey data recorded on days with stringency levels less than 40, 40-55, 55-70, and 70+, are shown in panels a, b, c, and d, respectively.}
    \label{fig:stringency_school}
\end{figure}

\begin{figure}
    \centering
    \includegraphics[width = \textwidth]{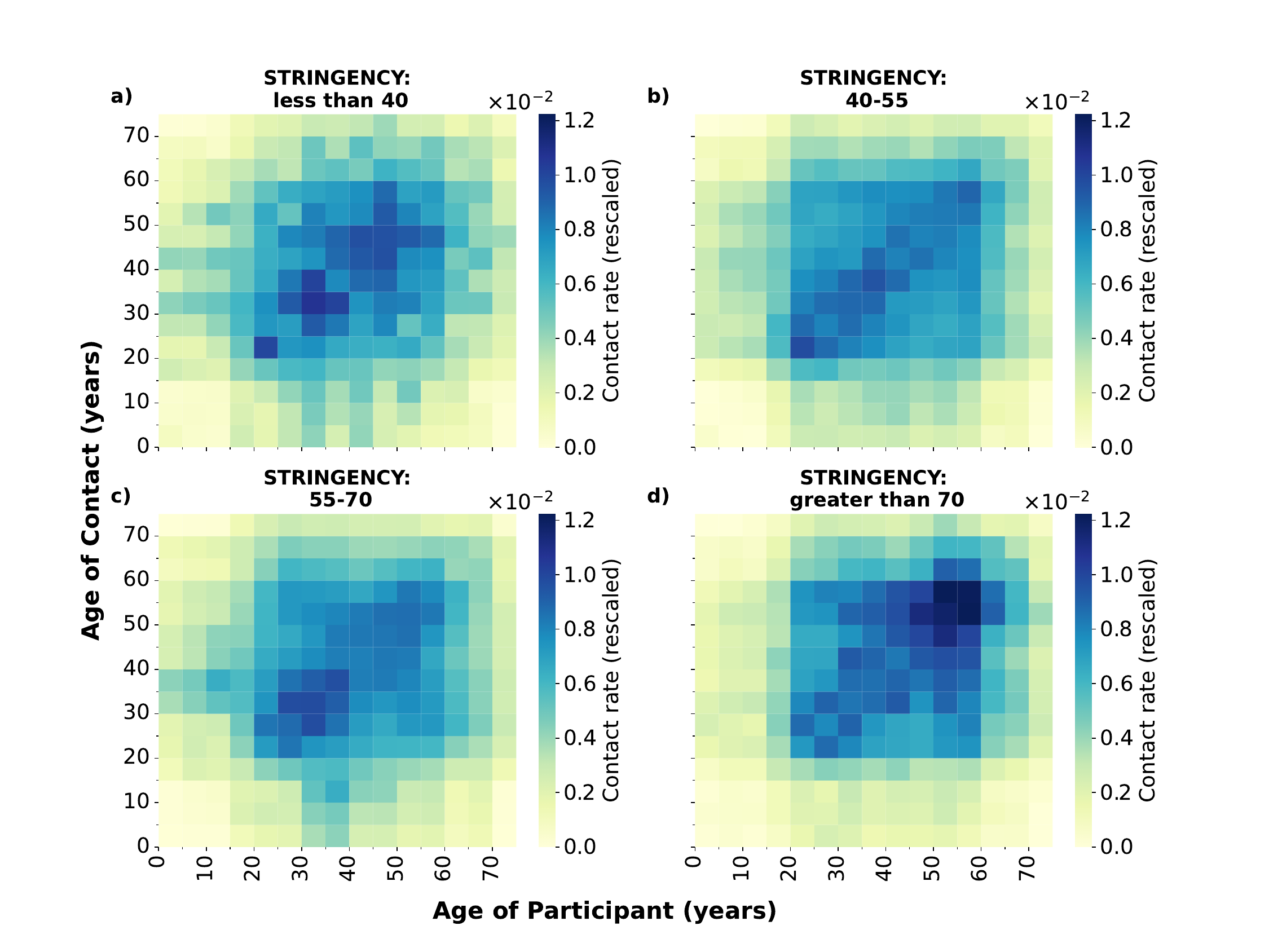}
    \caption{Workplace contact matrices derived from CoMix survey conducted in the United Kingdom, split on stringency of interventions on survey days. Stringency is quantified using a composite measure based on nine response indicators, such as school/work closures and travel bans. On a given day, the index takes a value between 0 and 100, with 100 indicating the strictest setting. Workplace contact patterns derived from survey data recorded on days with stringency levels less than 40, 40-55, 55-70, and 70+, are shown in panels a, b, c, and d, respectively.}
    \label{fig:stringency_work}
\end{figure}

We computed the overall and location-specific (household, school, workplace) contact patterns derived from survey days with stringency values in four different bands (less than 40, 40-55, 55-70, 70+). The overall patterns were similar for stringency values up to 70 (Figure~\ref{fig:stringency_overall}a-c). Above 70, contact patterns were more similar to the household contact patterns (Figure~\ref{fig:stringency_overall}d), likely reflecting the higher proportion of contacts occurring in households under more stringent policy settings (e.g., stay-at-home orders). The household, school and workplace contact patterns were largely similar across all stringency settings (Figure~\ref{fig:stringency_household}~-~\ref{fig:stringency_work}). 

\newpage\phantom{blabla}

\newpage\phantom{blabla}

\newpage\phantom{blabla}

\newpage\phantom{blabla}

\newpage
\section{Contact survey datasets used in analysis}
\label{surveys}
Survey datasets were sourced from a social contact data sharing initiative, where an online collection of social contact data and analysis methods for infectious disease modellers are available \cite{willem2020socrates}. The datasets used in this analysis are all publicly available \cite{joel_mossong_2020_3874557,carlos_g_grijalva_2020_3874805,alessia_melegaro_2020_3886638,guillaume_beraud_2020_3886590,kathy_leung_2020_3874808,zhang_2020_3878754,hens_niel_2020_4059864,willem_lander_2020_4302055,zhang_juanjuan_2022_7326686,backer_2022_7276465,gimma_2022_6542524,gimma_2022_6362899,gimma_2022_6362906,gimma_2022_6362898,gimma_2022_6362893,gimma_2022_6362888,gimma_2022_6362887,gimma_2022_6362865,gimma_2022_6362870,gimma_2022_6542668,gimma_2022_6542664,gimma_2022_6542657,gimma_2022_6535357,gimma_2022_6535344,gimma_2022_7257433,gimma_2022_6535313}.

\newpage
\section{Method for processing and comparing contact matrices}
\label{matrix_comparison}
\begin{figure}
    \centering
    \includegraphics[width = \textwidth]{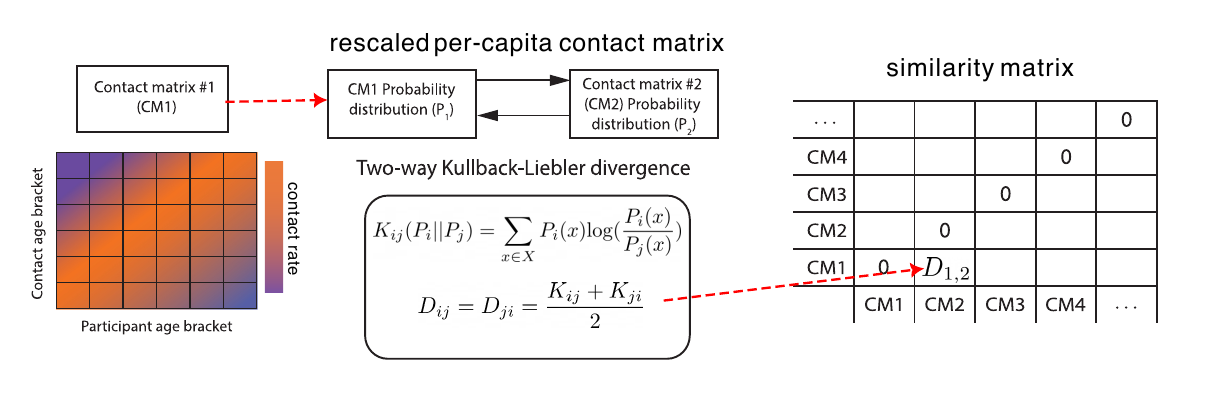}
    \caption{Schematic of contact matrix comparison. We rescale each contact matrix into a probability distribution (P), which allows us to compare two matrices by computing the two-way Kullback-Liebler (KL) divergence. We compare matrices derived from a large set of contact surveys to assess survey features that produce structural similarity.}
    \label{fig:schematic_comparison}
\end{figure}

To compare two per-capita contact matrices, we process each matrix into a probability distribution and compute the two-way Kullback-Liebler (KL) divergence (Figure \ref{fig:schematic_comparison}). Each per-capita contact matrix $c'_{nm}$ represents the contact rate between an individual in age group $n$ with an individual in age group $m$. To produce a probability distribution from each per-capita contact matrix, we divide each element of a contact matrix by the sum of the matrix. Re-scaling each matrix allows us to compute the KL divergence between two matrices to quantify similarity. The KL divergence ($K_{ij}$) captures the distance between a probability distribution ($P_i$) and a reference probability distribution ($P_j$) over some sample space. We compute the two-way KL divergence ($D_{ij}$), where we calculate the average of $K_{ij}$ and $K_{ji}$, between each pairwise combination of the re-scaled, per-capita contact matrices. From these similarity values, we construct a similarity matrix where each element $i$,$j$ is equal to $D_{ij}$, representing the similarity of contact matrices $i$ and $j$. This matrix is symmetric due to the reciprocal nature of the two-way KL divergence measure (i.e., $D_{ij} = D_{ji}$).

\newpage\phantom{blabla}

\newpage
\section{Methods for spectral clustering algorithm parameterisation \& assessing robustness}
\label{spec_cluster}
To assess robustness of our spectral clustering approach, we ran the clustering process $N_r=1000$ times for each combination of the number of clusters $N_c$ and number of neighbours $N_n$ considered. We record the resulting cluster assignment ($G^w$) computed in each iteration $w$:
\begin{equation}
    G^w = \{g_1,g_2,g_3,...,g_{33}\}\,, 
\end{equation}
where $g_x$ refers to the cluster assigned to survey $x$. Each cluster assignment is of size 33 (i.e., $|G^w| = 33$) for the 33 contact surveys analysed.

We define an adjacency matrix ($a^w$) for each cluster assignment $G^w$ that relates each survey to one another based on whether they were assigned to the same cluster:
\begin{equation}
    a^w_{ij} = \begin{cases} 
                      0, & \text{if $g_i \neq g_j$} \\
                      1, & \text{if $g_i = g_j$} 
             \end{cases} \,, 
\end{equation}

We assess robustness ($R$) over the set of adjacency matrices ($A$) produced from $N_r$ iterations of the clustering process using parameter values $N_c$ and $N_n$ by computing:
\begin{equation}
    R(N_c,N_n) = \frac{A'}{A''}\,, 
\end{equation}
\begin{equation}
    A' = \frac{\sum_{a \in A}\sum^{33}_{r=1}\sum^{33}_{v=1}a_{rv}}{N_r}\,, 
\end{equation}
\begin{equation}
    A'' = \sum^{33}_{r=1}\sum^{33}_{v=1}A^{T}_{rv}\,, 
\end{equation}
\begin{equation}
    A^{T}_{ij} = \begin{cases} 
                      0, & \text{if $a_{ij} \neq 1$ for all $a \in A$} \\
                      1, & \text{otherwise}
             \end{cases} \,, 
\end{equation}
where $A'$ is the mean number of edges in each adjacency matrix in A, and $A''$ is the number of edges in the union ($A^{T}$) of all the adjacency matrices in $A$.

$R$ must be between 0 and 1, with values of $R$ closer to 1 indicating more robustness across the cluster assignments. When $R=1$, this indicates all cluster assignments were identical over the entire set.

We computed $R(N_c,N_n)$ when $N_c$ was varied between 3 and 10, and $N_n$ varied between 4 and 8 (Table \ref{tab:robustness}). We found generally there are robust groupings achieved across most combinations of $N_c$ and $N_n$.
\setlength\arrayrulewidth{1pt}
\begin{table}[h]
\centering
\caption{Spectral clustering robustness ($R$) across different values for $N_c$ and $N_n$}
\begin{tabular}{|l|l|l|l|l|l|}
\hline
\textbf{\pmb{$N_c$ \textbackslash $N_n$}} & \textbf{$\pmb{N_n=4}$} & \textbf{$\pmb{N_n=5}$} & \textbf{$\pmb{N_n=6}$} & \textbf{$\pmb{N_n=7}$} & \textbf{$\pmb{N_n=8}$} \\ \hline
\textbf{$\pmb{N_c=3}$}                                    & 1.00                            & 1.00                            & 1.00                            & 1.00                            & 1.00                            \\ \hline
\textbf{$\pmb{N_c=4}$}                                    & 1.00                            & 1.00                            & 1.00                            & 1.00                            &                      1.00       \\ \hline
\textbf{$\pmb{N_c=5}$}                                    & 1.00                            & 1.00                            & 1.00                            & 1.00                            & 1.00                            \\ \hline
\textbf{$\pmb{N_c=6}$}                                    & 0.89                            & 1.00                            & 1.00                            & 1.00                            &                      1.00       \\ \hline
\textbf{$\pmb{N_c=7}$}                                    & 1.00                            & 1.00                            & 1.00                            & 1.00                            & 1.00                            \\ \hline
\textbf{$\pmb{N_c=8}$}                                    & 1.00                            & 1.00                            & 1.00                            & 1.00                            & 1.00                            \\ \hline
\textbf{$\pmb{N_c=9}$}                                    & 0.76                            & 0.89                            & 0.87                            & 0.87                            & 0.75                            \\ \hline
\textbf{$\pmb{N_c=10}$}                                   & 1.00                            & 1.00                            & 1.00                            & 0.84                            & 1.00                            \\ \hline
\end{tabular}
\label{tab:robustness}
\end{table}

To investigate if coherent groupings emerged using different numbers of clusters ($N_c$) in the spectral clustering algorithm, we performed spectral clustering on the survey set with values of $N_c$ varying between three and ten, and $N_n=5$ (Table \ref{tab:clustering}). Across all values considered for $N_c$, we found the POLYMOD and CoMix surveys were consistently separately grouped (i.e., no groups contained both POLYMOD and CoMix surveys). For example, at $N_c=10$, the CoMix surveys has been split among six separate clusters each made up entirely of CoMix surveys. Besides the POLYMOD and CoMix split discussed in the main text, there appeared to be no coherent groupings revealed among clusterings using higher $N_c$ values that could be explained by survey factors (e.g., country-level characteristics, such as geographic similarity).
\renewcommand{\tabcolsep}{1pt}
\begin{table}[]
\centering
\caption{Spectral clustering of contact patterns derived from different surveys using different values of $N_c$ and $N_n=5$. The cluster number assigned under each $N_c$ setting is shown alongside each survey. (Note the robustness of $N_c=9$ (Table \ref{tab:robustness}) when interpreting results for this setting)
}
\resizebox{\textwidth}{!}{%
\begin{tabular}{|c|c|c|c|c|c|c|c|c|}
\hline
\Large{\textbf{Country}} & \Large{\textbf{ $\pmb{N_c=3}$ }} & \Large{\textbf{ $\pmb{N_c=4}$ }} & \Large{\textbf{$\pmb{N_c=5}$}} & \Large{\textbf{$\pmb{N_c=6}$}} & \Large{\textbf{$\pmb{N_c=7}$}} & \Large{\textbf{$\pmb{N_c=8}$}} & \Large{\textbf{$\pmb{N_c=9}$}} & \Large{\textbf{$\pmb{N_c = 10}$}} \\ \hline
Austria (C) & 1 & 2 & 1 & 6 & 4 & 8 & 4 & 6 \\ \hline
Croatia (C) & 1 & 1 & 3 & 1 & 3 & 6 & 1 & 10 \\ \hline
Denmark (C) & 1 & 2 & 1 & 6 & 1 & 1 & 4 & 6 \\ \hline
Estonia (C) & 1 & 2 & 3 & 1 & 3 & 6 & 1 & 9 \\ \hline
Finland (C) & 1 & 1 & 5 & 5 & 5 & 4 & 7 & 10 \\ \hline
France (C) & 1 & 1 & 5 & 5 & 5 & 4 & 7 & 2 \\ \hline
Greece (C) & 1 & 2 & 3 & 1 & 4 & 8 & 4 & 4 \\ \hline
Hungary (C) & 1 & 2 & 3 & 1 & 4 & 8 & 4 & 6 \\ \hline
Italy (C) & 1 & 2 & 3 & 1 & 4 & 8 & 2 & 4 \\ \hline
Lithuania (C) & 1 & 2 & 1 & 6 & 1 & 1 & 5 & 8 \\ \hline
Netherlands (C) & 1 & 2 & 3 & 1 & 4 & 8 & 4 & 4 \\ \hline
Portugal (C) & 1 & 1 & 5 & 5 & 5 & 4 & 7 & 2 \\ \hline
Slovakia (C) & 1 & 2 & 3 & 1 & 3 & 6 & 2 & 9 \\ \hline
Slovenia (C) & 1 & 2 & 1 & 6 & 1 & 1 & 5 & 8 \\ \hline
Spain (C) & 1 & 1 & 5 & 5 & 5 & 4 & 7 & 2 \\ \hline
Switzerland (C) & 1 & 2 & 3 & 1 & 3 & 6 & 1 & 10 \\ \hline
United Kingdom (C) & 1 & 2 & 3 & 1 & 3 & 6 & 2 & 9 \\ \hline
China (Shanghai), 2018 & 2 & 3 & 2 & 2 & 2 & 3 & 6 & 1 \\ \hline
China (Wuhan), 2019 & 2 & 3 & 2 & 2 & 2 & 3 & 6 & 1 \\ \hline
Peru, 2011 & 2 & 3 & 2 & 4 & 7 & 5 & 9 & 3 \\ \hline
Zimbabwe, 2013 & 2 & 3 & 2 & 2 & 2 & 3 & 6 & 1 \\ \hline
Belgium (P) & 3 & 4 & 4 & 3 & 6 & 2 & 8 & 5 \\ \hline
Belgium, 2006 & 3 & 4 & 4 & 4 & 7 & 2 & 8 & 5 \\ \hline
Belgium, 2010 & 3 & 4 & 4 & 3 & 6 & 7 & 3 & 7 \\ \hline
Finland (P) & 3 & 4 & 4 & 3 & 6 & 7 & 3 & 7 \\ \hline
France, 2012 & 3 & 4 & 4 & 3 & 6 & 7 & 3 & 7 \\ \hline
Germany (P) & 3 & 4 & 4 & 3 & 6 & 7 & 3 & 7 \\ \hline
Hong Kong, 2016 & 3 & 4 & 4 & 4 & 7 & 5 & 9 & 3 \\ \hline
Italy (P) & 3 & 4 & 4 & 4 & 7 & 5 & 9 & 3 \\ \hline
Luxembourg (P) & 3 & 4 & 4 & 4 & 7 & 2 & 8 & 5 \\ \hline
Netherlands (P) & 3 & 4 & 4 & 4 & 7 & 5 & 9 & 3 \\ \hline
Poland (P) & 3 & 4 & 4 & 4 & 7 & 2 & 8 & 5 \\ \hline
United Kingdom (P) & 3 & 4 & 4 & 3 & 6 & 7 & 3 & 7 \\ \hline
\end{tabular}}
\label{tab:clustering}
\end{table}

 \end{document}